\documentclass[pre,twocolumn]{revtex4}

\usepackage{epsfig}

\usepackage{amssymb}
\usepackage{amsmath}
\usepackage[dvips]{color}
\usepackage{lscape}
\usepackage{longtable}
\usepackage{rotating}
\usepackage{float}

\usepackage{bm}

\newcommand{\be}{\begin{equation}}
\newcommand{\ee}{\end{equation}}
\newcommand{\bea}{\begin{eqnarray}}
\newcommand{\eea}{\end{eqnarray}}
\newcommand{\pat}{\partial}
\newcommand{\del}{\nabla}

\def\pmb#1{\setbox0=\hbox{#1}
        \kern-.025em\copy0\kern-\wd0
        \kern.05em\copy0\kern-\wd0
        \kern-.025em\raise.0433em\box0 } 
\def\div{\pmb {$\nabla$}}

\def\bnabla{\bm{\nabla}}

\begin{document}



\title{Heat transfer mechanisms in bubbly Rayleigh-B\'enard convection}
\author{Paolo Oresta$^1$, Roberto Verzicco$^2$, Detlef Lohse$^1$,
and Andrea Prosperetti$^{1,3}$}
\affiliation{
$^1$ Physics of Fluids Group, 
Department of Science and Technology, J.\ M.\
Burgers Centre for
Fluid Dynamics, and Impact Institute,  University of Twente, P.O. Box 217, 7500 AE Enschede, The Netherlands\\
$^2$ Department of Mechanical Engineering, University of Rome ``Tor Vergata'',
Via del Politecnico 1, 00133 Rome, Italy\\
$^3$ Department of Mechanical Engineering,
The Johns Hopkins University, Baltimore MD 21218, USA
}
\date{\today}





\date{\today}

\setcounter{page}{1}

\begin{abstract}
The heat transfer mechanism in Rayleigh-B\'enard convection in a 
liquid with a mean temperature close to its boiling point is studied through
numerical simulations with point-like vapor bubbles, which are allowed
to grow or shrink through evaporation and condensation and which act 
back on the flow both thermally and mechanically. 
It is shown that the effect of the bubbles is strongly 
dependent on the ratio of the sensible heat to the latent heat as embodied 
in the Jacob number $Ja$. For very small $Ja$ the bubbles 
stabilize the flow by absorbing heat in the warmer regions and releasing it 
in the colder regions. With an increase in $Ja$, the added 
buoyancy due to the bubble growth destabilizes the flow with respect to 
single-phase convection and considerably increases the Nusselt number. 
\end{abstract}

\maketitle


\section{Introduction}

Thermal convection is an omnipresent phenomenon in nature and technology.
The idealized version thereof is Rayleigh-B\'enard (RB) convection -- a 
single-phase fluid in a closed container heated from below and cooled from
above. A key question is the dependence of the heat transfer rate (as 
measured by the Nusselt number) for given
temperature difference between the hot bottom  and cold 
top  plate (i.e., Rayleigh number), 
given fluid (i.e., Prandtl number), and given aspect ratio. 
In the last two decades there has been tremendous progress on 
this and  related questions by experiment, theory and numerical 
simulation, see \cite{kad01} and \cite{ahl09} for a recent review.
Most of the work focused on RB convection for single-phase flow. Various 
situations in the process and energy industries, however, involve convection 
in the presence of phase change, e.g. condensing vapors and boiling liquids. 

The effectiveness of boiling as a heat transfer mechanism has been known for 
centuries and the process has formed the object of a very large number 
of studies \cite{dhi98}. Most of the focus has been on the process by 
which the high thermal resistance opposed by the visco-thermal layer 
adjacent to the hot surface is decreased by the vapor bubbles, the two 
main mechanisms believed to be micro-convection and latent heat transport. 
Another significant effect of the bubbles, however, is to promote strong  
convective currents in the liquid, thus helping remove the heated layer 
near the hot wall. This aspect of the process forms the object of the 
present study.

In an actual experiment all the processes occur at the same time and it 
is next to impossible to separately quantify their relative importance. 
Numerical simulation appears to be a promising tool for this purpose. 
Ideally, a simulation should be able to resolve individual bubbles and follow 
their evolution but, with the present capabilities, only so few bubbles 
can be simulated to this level of detail that it would be 
very difficult to draw conclusive results 
\cite{bun02a,bun02b,esm05,MukherjeeDhir04}.
Therefore one has to fall back on point-bubble models in which the 
interaction of the individual bubbles with the surrounding liquid is  
parameterized. This approach has proven  valuable in the study of 
turbulence in particle-laden flows (see e.g. 
\cite{Elg,Boi98,FerranteElghobashi03}), 
in liquids with 
gas -- rather than vapor -- bubbles \cite{cli99,maz03a,maz03b} and for 
Taylor-Couette flow with microbubbles inducing drag reduction \cite{sug08b}. 

Many important physical mechanisms have been elucidated by these means 
and one may therefore hope that similar insights might be  achieved by
extending this line of research accounting for phase change 
processes, and the accompanying bubble growth and collapse, in a similar 
way. Thus, to the fluid-mechanic bubble-liquid interaction model used  
in our earlier work (\cite{maz03a,maz03b}), we add here models for the heat 
transfer and phase change between the bubbles and the surrounding liquid 
along the lines of Refs.\ \cite{leg98b,iva04}.

\begin{figure}
\centering
\includegraphics[width=7cm,angle=0]{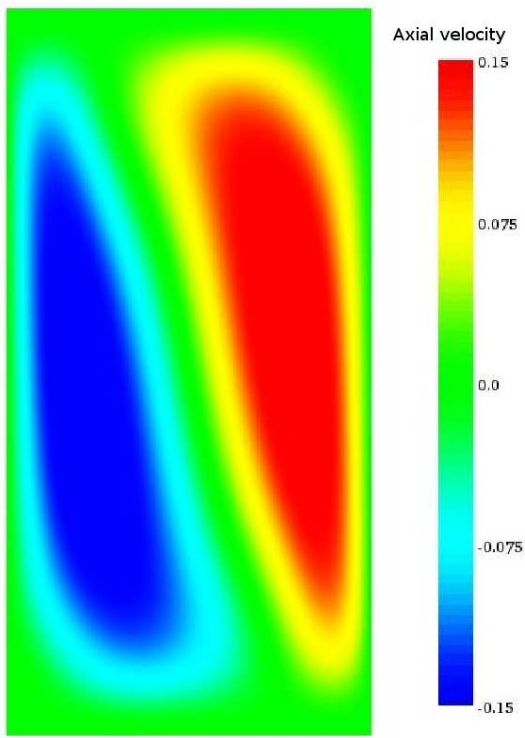}
\caption{Vertical velocity in the plane of symmetry of the full
 cylinder 
in the absence of bubbles;  $Ra\,=\,2\times 10^5$, $Pr$ = 1.75, $Nu$ = 4.75.
As throughout the paper, the  velocity is 
made dimensionless by using the free-fall velocity 
$(\beta gH (T_h-T_c))^{1/2}$. 
}
\label{fig:snglphs}
\end{figure}

The standard single-phase RB convection under the Boussinesq approximation 
is controlled by the Rayleigh number
\be
            Ra=\frac{g \beta (T_h-T_c) H^3}{\nu \kappa},
\label{defra}
\ee
where $T_h$ and $T_c$ are the temperatures of the hot (bottom) and 
cold (top) plate, respectively,
$H$ is the height of the convection cylinder, 
$g$ the gravitational acceleration,
$\kappa$ the thermal diffusivity of the liquid,
$\nu$ its kinematic viscosity,
and $\beta$ the isobaric thermal expansion coefficient). 
The Prandtl number is defined as
\be
    Pr\,=\,\frac{\nu}{\kappa}
\ee
and the aspect ratio of the cylinder  as the ratio of the diameter to 
the height. In this paper we consider  convection for which, without bubbles, 
$Ra\,=\,2\times 10^5$ and  $Pr$ = 1.75 (water at $100^\circ C$); the aspect 
ratio is 1/2 and the cell is cylindrical. With these parameter values, in the 
absence of bubbles, there is a convection roll with fluid rising along one 
side of the cell and descending along the opposite side 
(see figure~\ref{fig:snglphs}); the Nusselt number has the value 4.75. 

Vapor bubbles introduce a crucial new parameter, the Jacob number
\be
 Ja\,=\, \frac{\rho c_p(T_h-T_{sat})}{\rho_V L} \, 
\label{defja}
\ee
in which $L$ is the latent heat, $\rho_V$ and $\rho$ the vapor and liquid 
density, respectively, $c_p$ the liquid specific heat and  $T_{sat}$ the 
saturation temperature of the liquid. With the parameter values used in this 
study, hydrostatic pressure variations are not sufficient to cause a 
significant change of $T_{sat}$, which therefore is taken as a 
constant equal to the average of the hot and cold plate temperatures. 
Physically, $Ja$ represents the ratio of the sensible heat to the latent heat.
A very small Jacob number may be thought of as a very large value of the 
latent heat, which will tend to limit the volume change of the bubbles due to 
evaporation or condensation. 

For $Ja=0$ the latent heat is effectively infinite and bubbles cannot grow or 
shrink; they maintain their initial diameter at nucleation, which we take to 
be 25 $\mu$m. Another control parameter in our model  
is the total number $N_b$ of bubbles in the cylinder. Though in real systems 
this number will fluctuate in time somewhat, here we take it as constant: 
Whenever a bubble reaches the top of the cylinder and is removed, 
a new bubble of the standard initial size ($25\,\,\mu$m) is nucleated at the 
bottom plate at some random position. 


\section{Model}
\label{sec:model}

We study the problem in the standard Boussinesq approximation augmented 
by the momentum and energy effects of the bubbles, treated as points. 
When the volume occupied by the bubbles is very small, the liquid continuity 
equation retains the standard incompressible form 
\be 
   \bnabla \cdot {\bf u}\,=\,0
\ee
in which {\bf u} is the liquid velocity field. The momentum equations is 
\be
 \rho {D {\bf u}\over D t} 
\,=\,-\div p + \mu\del^2{\bf u} +\beta \rho (T-T_{sat}){\bf g}+
\sum_n{\bf f}_n\delta({\bf x}-{\bf x}_n)
\label{momeq}
\ee
where $D/Dt$ is the convective derivative, $p$ and $T$ are the pressure and 
temperature, and $\mu = \nu \rho$ the dynamic viscosity. The effect of the 
bubbles has been approximated  in the standard way by modeling
them as point-like 
sources of momentum, which is adequate when the volume fraction is small and 
the bubble radius is smaller than the fluid length scales 
(see e.g. \cite{maz03b}).

The position of the $n$-th bubble is denoted by ${\bf x}_n$ and the force 
${\bf f}_n$ that it applies on the liquid is modelled as 
(see e.g.\ \cite{max94,maz03b}) 
\be
  {\bf f}_n\,=\, {4\over 3}\pi R_{bn}^3\rho\left(\left. 
{D{\bf u}\over Dt}\right|_{{\bf x}_n}-{\bf g}
\right) 
\label{forcbub}
\ee
in which $R_{bn}$ is the radius of the $n$-th bubble and the liquid 
acceleration is evaluated at the position of the bubble. 
A similar term multiplied by the vapor, rather than the liquid, density has 
been neglected here. 

The liquid energy equation takes the form 
\be
 \rho c_p {D T\over Dt} \,=\,k\del^2 T + \sum_n Q_n\delta({\bf x}-{\bf x}_n)
\label{liqen}
\ee
where $k = \kappa \rho c_p$ is the liquid  thermal conductivity 
and $Q_n$ is the energy source or sink due to phase change of the $n$-th 
bubble. 
We model the thermal exchange between the $n$-th bubble and the liquid 
by means of a heat transfer coefficient $h_{bn}$ and write 
\be
Q_n= 4\pi R_{bi}^2 h_{bn}(T_{sat}-T_n) 
\label{modqi}
\ee
where  $T_n=T({\bf x}_n,t)$ is the liquid temperature evaluated at the 
position of the $n$-th bubble. In writing this relation we have used the 
fact that, for moderate temperature differences, phase change is 
slow and the bubble surface remains essentially at saturated conditions 
(see e.g. \cite{YangProsperetti08}). 

The expressions (\ref{forcbub}) and (\ref{modqi}) and the use of point 
sources of momentum in (\ref{momeq}) and of energy in (\ref{liqen}) 
assume that the bubbles interact only through the average fields but not 
directly, which is a reasonable approximation at the vapor volume fractions 
considered here (see e.g.\ \cite{maz03a,maz03b}). 

Part of the system energy is carried by the bubble phase. If 
$E_b$ denotes the energy of a single bubble, $n$ the bubble number 
density and ${\bf v}$ the bubble velocity, conservation of this component 
of the system energy is expressed by 
\be
 {\pat \over \pat t}(nE_b)+\div \cdot (nE_b{\bf v})\,=\, 
-\sum_n Q_n\delta({\bf x}-{\bf x}_n)
\label{buben}
\ee
where the small $p\,dV$ contribution has been neglected. 
Adding (\ref{liqen}) and (\ref{buben}) gives an equation for the balance of 
the total system energy, namely 
\be
 {\pat \over \pat t}[\rho c_p (T-T_{sat})+nE_b]
+\div \cdot [\rho c_p (T-T_{sat}) {\bf u}+
nE_b{\bf v}) ] \,=\, k \del^2 T.
\label{toten}
\ee

With the neglect of the vapor mass, the equation of motion for each bubble 
balances added mass, lift, and buoyancy, 
\bea
&& C_A\rho \left[{4\over 3}\pi R_b^3\left({D {\bf u} \over D t}-
{d{\bf v}\over dt}\right)+({\bf u}-{\bf v})
{d \over dt}\left({4\over 3}\pi R_b^3\right)\right]\nonumber \\ &&
\qquad -{1\over 2}\pi C_D \rho
R_b^2 |{\bf v}-{\bf u}|({\bf v}-{\bf u})
+{4\over 3}\pi R_b^3\rho {D {\bf u}\over D t}  \nonumber  \\ && 
\qquad \qquad +C_L{4\over 3}\pi R_b^3 
\rho\left(\div \times {\bf u}
\right)\times ({\bf v}-{\bf u})\nonumber \\ && \qquad 
-{4\over 3}\pi R_b^3\rho{\bf g}
\,=\,0
\label{bubeqmot}
\eea
in which $C_A$, $C_L$, and $C_D$ are the added mass, lift and drag 
coefficients, respectively. 
The uncertainty with which many of the terms of this equation are known 
is well appreciated in the literature (see e.g.\ \cite{mag00} or our
own work \cite{nie07}). 
Moreover, due to interaction with the wake, there might be history forces 
which have been neglected in (\ref{bubeqmot}) \cite{mei94,leg98,mag98}.  
Nevertheless, as written, the equation captures the basic effects of 
drag, buoyancy, and added mass which dominate the bubble-liquid interaction. 
After some rearrangement, the equation becomes 
\bea && 
 C_A{d{\bf v}\over dt}\,=\,(1+C_A){D{\bf u}\over Dt} - {3C_A\over R_b}
({\bf v}-{\bf u}){dR_b\over dt}\nonumber 
\\ && -{3\over 8}{C_D\over R_b}
 |{\bf v}-{\bf u}|({\bf v}-{\bf u}) -{\bf g} \nonumber \\ && 
+ C_L \left(\div \times {\bf u}
\right)\times ({\bf v}-{\bf u})
\eea
The bubble radius $R_b$ is calculated by balancing the latent heat 
associated to evaporation or condensation with the heat exchanged with 
the liquid 
\be
 L{d\over dt}\left({4\over 3}\pi R_b^3\rho_V\right)\,=\, 
- Q_n \, = \,  
4\pi R_b^2 h_b
(T-T_{sat}).
\ee
Since the 
bubble is assumed to be at saturation, $\rho_V$ is a constant and this 
equation can be simplified to the form 
\be
 {dR_b\over dt}\,=\, {h_b\over L \rho_V}(T-T_{sat})
\label{eqfrb}
\ee
in which $\rho_V\,=\,\rho_V(T_{sat})$.

In order to complete the mathematical formulation of the problem, definite 
choices must be made for several quantities. Since our bubbles are 
small and therefore will not deform very much,  we take $C_A=1/2$, 
the standard potential-flow value for a sphere (see e.g. \cite{bat67}), 
independent of the Reynolds number and of non-uniformities of the flow 
\cite{mag95,aut87,riv91,cha95}. 
The inviscid calculation of 
\cite{aut87} gives the same value for the lift coefficient; this value 
appears to be a reasonable estimate even at low to moderate Reynolds number 
(see figure 17 of \cite{leg98}). 
We model the drag coefficient as suggested by \cite{mei92,mei94},
\be
 C_D\,=\, {16\over Re_b}\left[ 1 +{Re_b \over 
8 +{1\over 2}(Re_b+3.315\sqrt{Re_b})}\right] 
\ee
in which $Re_b =2R_b |{\bf v}-{\bf u}|/\nu$ is the bubble Reynolds 
number. 

\begin{figure}
\centering
\includegraphics[width=8cm,angle=0]{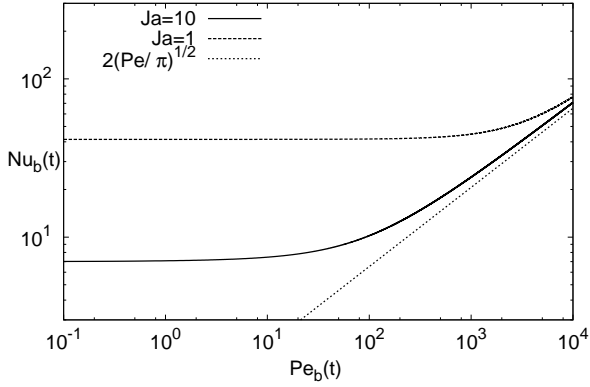}
\caption{The interpolation (\ref{Nufin}) for the 
dependence of the single-bubble Nusselt number on the P\'{e}clet 
number.}
\label{fig:NuvsPe}
\end{figure}

We express the heat transfer coefficient $h_b$ in terms of a single-bubble 
Nusselt number 
\be
    Nu_b\,=\,{2R_b h_b\over k}
\label{defnu}
\ee 
The dependence of $Nu_b$ on the parameters of the problem is complicated 
and has been studied by several authors (see e.g.\ 
\cite{leg98b,iva04}). 
In order to make progress we are forced to introduce some simplifications. 
The analysis of \cite{leg98b} shows that, as a function of the P\'{e}clet 
number
\be
 Pe_b\,=\,{2R_b|{\bf v}-{\bf u}|\over \kappa},
\ee
there are essentially two regimes. At low $Pe_b$, $Nu_b$ is approximately 
independent of $Pe_b$ and only depends on the Jacob number (\ref{defja}).
We call this value $Nu_{b,0}$
The functional relationship $Nu_{b,0}(Ja)$ in this regime has been variously 
parameterized by different authors. Reference \cite{leg98b} proposes a 
general form 
\be
  Nu_{b,0}\,=\, {16 \over \pi}Ja\, f(Ja)
\label{formnu0}
\ee
For the function $f(Ja)$ Ref. \cite{lab64} (corroborated by the more recent 
results of Ref.\cite{iva04}) proposes 
\be
  f(Ja)=\frac{\pi}{8Ja}+\frac{(6\pi^2)^{1/3}}{16}\frac{1}{Ja^{2/3}}+\frac{3}{4}
\label{eq:LabufJa}
\ee 
with which (\ref{formnu0}) becomes 
\be
   Nu_{b,0}\,=\,2+\left(\frac{6Ja}{\pi}\right)^{1/3}+\frac{12}{\pi}Ja
\label{eq:Nu0Lab}
\ee
For very 
large P\'{e}clet numbers, heat transfer is dominated by convection and 
the result is \cite{ruc59} 
\be
   Nu_{b,\infty}=2\sqrt{\frac{Pe_b}{\pi}}
\label{eq:Nupf}
\ee
We combine these two asymptotic forms in a way that smoothly interpolates 
between them:
\be
Nu_b \,=\, Nu_{b,0}\, 
\left[1+\left(\frac{Pe_b}{Pe_c}\right)^{n/2}\right]^{1/n} 
\label{Nufin}
\ee 
where $n\,\simeq\,2.65$ is determined by fitting the results of Refs.\
 \cite{ruc59} and \cite{iva04} and the crossover P\'eclet number $Pe_c$, 
defined by $Nu_{b,\infty} \,=\,Nu_{b,0}$, is $Pe_c= \pi Nu_{b,0}^2/4$.
The relation (\ref{Nufin}) is shown as a function of $Pe_b$ for $Ja$ = 
1 and 10 in figure \ref{fig:NuvsPe}. These results can be compared with 
the corresponding ones presented in figure 3 of \cite{leg98b} and are 
seen to provide an accurate representation of them. 

\section{Nusselt number}\label{sec:nu}

If the total energy equation (\ref{toten}) is averaged over time and 
integrated over the cylinder volume we find 
\be
\left.  \langle nE_b v_3- k \pat_3 T\rangle_{A,t}\right|_{z=H}
\,=\, \left. \langle  nE_b v_3- k \pat_3 T\rangle_{A,t}\right|_{z=0}
\label{totenfl}
\ee
where the subscript 3 denotes the vertical direction and 
$\langle \ldots \rangle_{A,t}$ the time and area average. In deriving this 
relation we have used the no-slip condition for the liquid phase and the 
assumed adiabaticity of the lateral walls. The bubble velocity on the bottom 
and top plates must account for the injection and removal of bubbles and 
therefore cannot be taken to vanish. 
A similar treatment of the bubble energy equation (\ref{buben}) gives 
\be
\left.  \langle nE_b v_3\rangle_{A,t}\right|_{z=H} -\left. 
\langle  nE_b v_3\rangle_{A,t}\right|_{z=0} \,=\, -{1\over \pi R^2}\left< 
\sum_n Q_n\right>_t
\label{bubenfl}
\ee
where $R$ is the radius of the cylinder. 
The summation in the last 
term is over 
all the bubbles contained in the system and the average is over time. 
Using (\ref{bubenfl}), (\ref{totenfl}) can equivalently be written as 
\be
\left.  -\langle k \pat_3 T\rangle_{A,t}\right|_{z=H} + \left.
 k \langle \pat_3 T\rangle_{A,t}\right|_{z=0} \,=\,{1\over \pi R^2}\left< 
\sum_n Q_n\right>_t 
\label{totenfl2}
\ee
which expresses the obvious fact that any difference between the heat 
conducted out of the bottom plate and into the top plate is due to the energy 
stored in the bubbles. 

In single-phase natural convection the conventional definition of the 
Nusselt numbers $Nu_c $ and $Nu_h $ at the hot and cold plates is
\be
  Nu_{c,h} \,=\,- {H \over \Delta}\left. \langle \pat_3 T \rangle_{A,t}
\right|_{z=H,z=0} 
\label{sphnu}
\ee 
In the single-phase case this quantity may be considered as a 
total dimensionless heat flux, but this interpretation would be 
incorrect here as it disregards the effect of the bubbles. Here 
the proper quantity to be regarded as the total dimensionless heat flux 
would be 
\be 
N^*_{c,h}\,=\, {H\over k \Delta} \left.
\langle nE_b v_3- k \pat_3 T\rangle_{A,t}\right|_{z=H,0}
\ee 
which, by (\ref{totenfl}), satisfies 
\be 
N^*_{h}\,=\, N^*_{c}
\ee
as expected. 
However, since the point of this paper is to show the impact of the bubbles 
on what would be considered the heat flux in single-phase convection, it 
is preferable to present our results in terms of $Nu_{h,c}$ rather than 
$N^*_{h,c}$. 

The definitions (\ref{sphnu}) lead to 
\be
 Nu_c -Nu_h \,=\, {H\over \pi R^2 k\Delta }\left< 
\sum_n Q_n\right>_{t}. 
\label{nusdif}
\ee
Separate expressions for $Nu_c $ and $Nu_h $ can be found by 
using another relation which can be derived by multiplying (\ref{liqen}) by 
$z-{1\over 2}H$ and integrating over the volume of the cylinder with the result 
\bea &&
\overline{Nu}\,\equiv\, {1\over 2}\left(Nu_c +Nu_h \right)\,=\,
1+{H \over \kappa\Delta}
\langle u_3(T-T_{sat}) \rangle_{V,t}\nonumber \\ && \qquad +
{1\over \pi R^2 k \Delta}\left< \sum_n \left(z_n-{1\over 2}H 
\right)Q_n \right>_t 
\label{avenus}
\eea
in which $\langle \ldots \rangle_{V,t}$ denotes a time and volume average; 
in the following we refer to $\overline{Nu}$ as the average Nusselt number. 
By using this relation and  (\ref{nusdif}) we have 
\be
 Nu_h \,=\,1+{H \over \kappa \Delta}\langle u_3(T-T_{sat})
\rangle_{V,t} 
+ {1\over \pi R^2 k \Delta}
\left< \sum_n \left(z_n-H \right)Q_n \right>_t
\label{nubot}
\ee
and 
\be
 Nu_c \,=\,1+{H \over \kappa \Delta}\langle u_3 (T-T_{sat})
\rangle_{V,t} + {1\over \pi R^2 k \Delta}\left< \sum_n z_nQ_n \right>_t
\label{nutop}
\ee

The dimensionless heat fluxes $N^*_{h,c}$ can be reconstructed by noting 
that, since bubbles are injected with a small velocity and a small radius, 
the first term in the right-hand side of (\ref{bubenfl}) is much smaller 
than the second one and therefore, approximately, 
\be
 \left. \langle  nE_b v_3\rangle_{A,t}\right|_{z=H} \,\simeq\, 
-{1\over \pi R^2}\left< \sum_n Q_n\right>_t
\label{bubenflap}
\ee
Thus, at the hot plate,  
\be
N^*_h\, \simeq \, Nu_h  
\ee
and, at the cold plate, 
\be
N^*_c\, \simeq \, Nu_c + {H\over \pi R^2 k \Delta}
\left< \sum_n Q_n \right>_t \,=\,Nu_c +(Nu_c -Nu_h )
\ee
in the last step of which use has been made of (\ref{nusdif}). 

Just as the Nusselt number, 
the expressions for the kinetic and thermal dissipations $\epsilon_u$
and $\epsilon_\theta$ of standard single-phase natural convection are also 
affected by the bubble contribution to the liquid energy equation. These 
modified expressions are derived in the Appendix.

\section{Numerical methods}


Equations (\ref{momeq}) and (\ref{liqen}) have been written in cylindrical 
coordinates and discretized using staggered second-order-accurate finite 
difference schemes. The resulting algebraic system is solved by a 
fractional step method with the advective terms treated explicitly and the 
viscous terms computed implicitly by an approximate factorization technique 
(see \cite{VeOr} for details).  
The Poisson equation that enforces the flow incompressibility
is solved by a direct procedure which relies on trigonometric expansions in
the azimuthal direction and the FISHPACK package \cite{swartz} for the
radial and axial directions for which, therefore, a non-uniform mesh 
distribution can be used. 
The grid is non-uniform in the radial and axial directions and 
clustered towards the boundaries to adequately resolve the viscous 
and thermal layers. Following Verzicco and Camussi \cite{Verz3}, we used a 
grid with $33\times 25\times 65$ points, respectively, in  the azimuthal, 
radial and axial directions after having 
verified that this resolution is sufficient for the present Rayleigh and 
Prandtl numbers. 

Although the code can handle high-order multistep schemes, 
the time advancement of the solution has been carried out by a
simple second-order Adams-Bashforth procedure. For this problem,
the most severe limitation on the time step size is imposed by the 
bubble relaxation time which, especially for the smallest bubbles, 
is much more stringent than the flow stability condition.

The only relevant change with respect to the method described in \cite{Verz3}
is the presence of bubble-induced momentum and thermal forcings in the 
governing equations. 
The forcing due to each bubble is located at its center and therefore, 
when  (\ref{momeq}) and (\ref{liqen}) are discretized, it has to be 
replaced with an equivalent system of forcings at the grid nodes. 
For this purpose, since in a staggered grid arrangement the momentum cells 
in the three directions are all different, the force (\ref{forcbub}) exerted 
by the bubble is first decomposed into its radial, azimuthal and vertical 
components. Each one of these components is then distributed by suitable 
weighing among the 8 vertices of the surrounding momentum cell in the same 
direction. For example, for a radial force component $f$ at a position 
$r_i+ \xi \Delta r$, $\theta_j+\eta \Delta \theta$, $z_k+\zeta \Delta z$, 
with $\Delta r$, $\Delta \theta$ and $\Delta z$ the grid spacings and 
$0\leq \xi,\eta,\zeta <1$, the portion attributed to the node 
$(r_i,\theta_j,z_k)$ is 
\be
f(1-\xi)(1-\eta)(1-\zeta).
\ee
The system of 8 forces thus obtained produces the 
same net resultant and couple as the original bubble force.
The same strategy has been used for the thermal 
forcing so that the total amount of heat that each bubble exchanges with the 
liquid is preserved.

The bubble trajectory is computed using the Adams-Bashforth scheme 
for position and the Crank-Nicholson scheme for velocity. This latter 
implicit scheme avoids the numerical instability induced by the fast 
dynamics of the smallest bubbles. 
Equation (\ref{eqfrb}) for the bubble radius is integrated explicitly. 

\begin{figure}
\begin{center}
\includegraphics[width=8cm,angle=0]{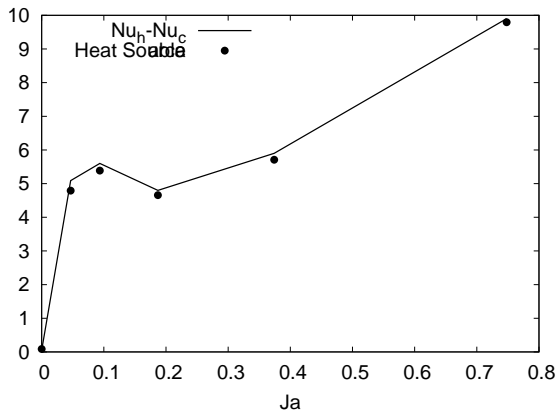}
\caption{The line shows the numerical results for the left-hand side 
of  (\ref{nusdif}) and the points those for the right-hand side. Equality of 
these two quantities is a stringent of the accuracy of the computation.
$N_b=5000$.
}
\label{fig:compsou}
\end{center}
\end{figure}

The numerical solver has been validated by monitoring the temporal evolution 
of a single bubble in a quiescent flow without a thermal 
field. 
Furthermore, our results have been 
compared with the theoretical prediction of the lateral force induced on 
a spherical bubble rising with a constant velocity in a viscous fluid near 
a vertical cylindrical wall. We followed the theoretical method 
of ref. \cite{Shi79} using the  free-slip 
boundary condition for the bubble surface instead of the no-slip condition 
used for a rigid particle. Another test of the numerical method and its 
implementation is offered by a comparison of the numerical results for the 
two sides of (\ref{nusdif}). Such a comparison is shown for a typical case 
in Fig.~\ref{fig:compsou}. 

\section{Implementation}
\label{impl}

From the numerical point of view, a significant practical difficulty of the 
present problem is the large difference between the flow time scale and the 
times over which bubbles grow and collapse. In order to have reasonable 
execution times of our computer code it has been necessary to limit this 
difference by adopting a small cylinder
 size; we have taken a height $H=17.9$ mm 
and a diameter $2R=8.94$ mm. Furthermore, 
in order to limit the number of spatial cells 
necessary to resolve the flow it is necessary to limit the Rayleigh 
number, which can be achieved by taking a small temperature difference; 
we take $T_h-T_c= 0.25$ K. With these values and the physical 
properties of water at 373 K, we have $Ra=2\times 10^5$. Since 
$T_{sat}={1\over 2}(T_h+T_c)$, the hot plate is 0.125 K hotter than 
the saturation temperature, which in reality would not be a superheat 
sufficient to nucleate bubbles. This is another respect in which our model 
deviates from reality. On the other hand, since our focus here is the 
bubble effect on the thermal convection, rather than the actual heat 
removal from the plate due to bubble formation, the compromise that is 
forced on us is less damaging than it might be in a study of boiling heat 
transfer. 

The calculation is started without bubbles and run until the steady state 
shown in figure ~\ref{fig:snglphs} is 
reached. At this point 25 $\mu$m-diameter bubbles are introduced randomly 
throughout the volume of the cylinder attributing to each one the local liquid 
velocity. From this point on, whenever a bubble reaches the top plate, it is 
removed and a new 25$\mu m$-diameter 
bubble is introduced at a random position on the bottom 
plate. The new bubble is placed at a height above the plate equal to its 
radius  and it is given the local liquid velocity. Bubbles reaching 
the lateral vertical wall of the cylinder are assumed to bounce elastically. 

In order to avoid possible numerical problems due to the disappearance or 
excessive growth of bubbles, we have imposed artificial limits on the minimum 
and maximum bubble diameters equal to 0.82 $\mu$m and 258 $\mu$m respectively. 
We found however that 
these limits are never approached in our simulations. Since bubbles never 
condense completely, the total number of bubbles is constant in time. 

\begin{figure}
\begin{center}
\includegraphics[width=8cm,angle=0]{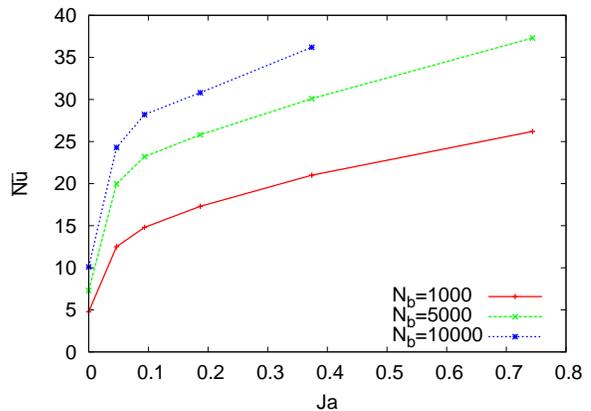}
\caption{Averaged Nusselt number $\bar Nu$ 
vs Jakob number for three different numbers of bubbles. 
}
\label{fig:numed}
\end{center}
\end{figure}

\section{Results}\label{sec:results}

Since bubbles tend to grow in volume in hotter liquid region, thus aiding 
buoyancy, and to condense in colder regions, they have a de-stabilizing 
effect on natural convection. These effects are clearly the stronger the 
larger the volume change. As explained before, in the present model this 
feature can be 
controlled by controlling the Jacob number (\ref{defja}). A very small Jacob 
number may be thought of as a very large latent heat, which 
will tend to limit the volume change of the bubble, while, conversely, a 
large Jacob number would enhance the destabilizing effect. 

While this is the major effect, there are other minor ones which operate 
in the opposite direction. For example, bubbles in a hot liquid region, for 
which $T>T_{sat}$, will tend to cool the liquid by absorbing heat, and 
conversely in a colder liquid region. If $Ja$ is very small so that the 
bubble is prevented from growing appreciably, this process tends to 
eliminate the very temperature differences which drive the natural convection 
in the first place. All other things being equal, the break-even point between 
increased buoyancy due to bubble expansion and decreased liquid buoyancy 
due to the bubble-induced cooling will be for that value of the 
Jacob number at which the thermal expansion of the bubble equals the 
added weight of the liquid due to the increased density. It will be seen 
from our results that this balance occurs for very small $Ja$ so that, in 
most practical situations, the balance will tip in favor of the enhanced 
buoyancy effect.

\begin{figure}
\begin{center}
\includegraphics[width=8cm]{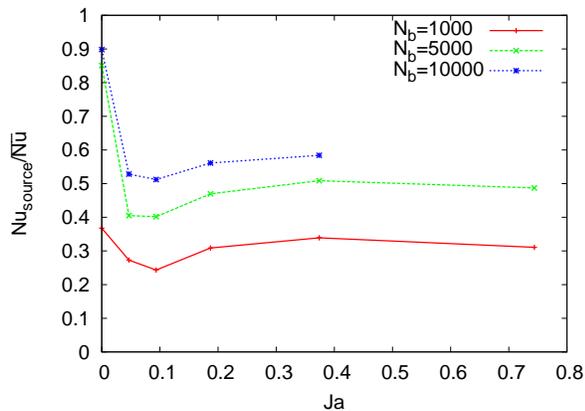}
\caption{Ratio of the bubble source term (\ref{defnus}) to the average 
Nusselt number (\ref{avenus}). 
At small $Ja$, the additional bubble-induced buoyant forcing is the 
dominant effect, while at large $Ja$ the bubbles act as direct carriers
of heat from the bottom to the top.
} 
\label{fig:nusource}
\end{center}
\end{figure}

Figure \ref{fig:numed} shows the effect on the average Nusselt number 
$\overline{Nu} = {1\over 2}(Nu_h +Nu_c )$, defined in (\ref{avenus}), of 
adding 1,000, 5,000 and 10,000 
bubbles to the basic single-phase RB flow; here, as in all the results shown, 
the Rayleigh number is $Ra\,=\,2\times 10^5$ and 
$Pr$ = 1.75. Figure \ref{fig:nusource} shows the fraction of the bubble 
contribution 
\be
Nu_{source} = {1\over\pi R^2 k \Delta}\left< \sum_n \left( z_n - {1\over 2}
H\right) Q_n \right>_t
\label{defnus}
\ee
to the average Nusselt number $\overline{Nu}$. 
The remaining fraction of the Nusselt number is accounted for by conduction 
and pure convection, i.e. the first two 
terms in the right-hand sides of (\ref{nubot}) and (\ref{nutop}). In both
figures the horizontal axis is the Jacob number, which we use as a control 
parameter to investigate the effect of the added bubble buoyancy. 

For $Ja$ = 0 the bubbles maintain their initial diameter at injection at the 
plate (25 $\mu$m) but, because they are kept at $T_{sat}$, they cool the 
hotter liquid regions and heat up the cooler ones. As noted before, this 
behavior tends to stabilize the RB convection and is responsible for the 
fact that, while in the absence of bubbles the flow consists of an annular 
roll with an approximately horizontal axis (Fig.~\ref{fig:snglphs}), the 
addition of $Ja=0$ bubbles changes it to a toroidal roll with a vertical 
axis. Because of this stabilizing effect, the cooling/heating 
operated by the bubbles accounts for a large fraction of the total heat 
transported and, indeed, it can be seen from Fig.~\ref{fig:nusource} that 
the bubble contribution (\ref{defnus}) is very large, up to about 90\% of 
the total for the 10,000 bubble case.

\begin{figure}
\begin{center}
\includegraphics[width=8cm,angle=0]{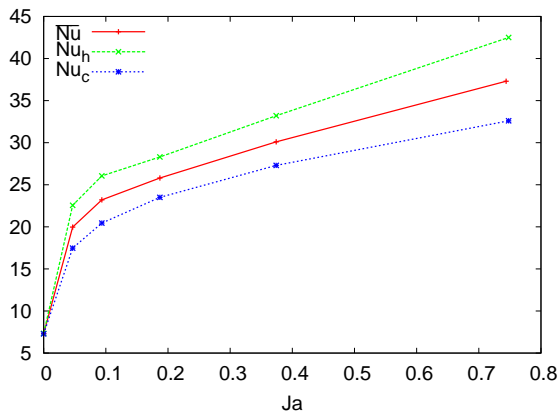}
\caption{Comparison between the Nusselt number computed at the top 
and at the bottom boundaries for 5,000 bubbles; the middle line is the 
average Nuselt number $\overline{Nu}={1\over 2}(Nu_h+Nu_c)$. 
Here we took $N_b=5000$.}
\label{fig:nuTBN2}
\end{center}
\end{figure}

As $Ja$ is increased, the Nusselt number increases very rapidly at first 
(Fig.~\ref{fig:numed}) due to the increased convection caused by buoyancy. 
As a consequence, the fraction of the total Nusselt number due to the bubbles 
(Fig.~\ref{fig:nusource}) undergoes a steep decline. With further increases 
of $Ja$, the Nusselt number keeps growing but at a more moderate rate. 
The minimum around $Ja$ = 0.1 observed in figure  \ref{fig:nusource} is due 
to a change of the flow structure as described later. 

Figure \ref{fig:nuTBN2} shows the Nusselt numbers computed at the top and 
bottom of the cylinder and their average for 5,000 bubbles; the behavior 
for the other bubble numbers is very similar. As shown by (\ref{nusdif}), the 
difference  $Nu_c -Nu_h $ is due to the heat exchanged  
between the bubbles and the liquid. As the Jacob number begins to increase, 
the energy absorbed by each bubble per unit time increases because of a direct 
increase in the heat transfer coefficient of each individual bubble 
(see Eq. \ref{eq:Nu0Lab}), and an increase in the convective component of 
the bubble heat flux caused by the faster rise velocity of a larger bubble 
(Eq.~\ref{eq:Nupf}). The moderation in the rate of growth of $Nu $ 
at larger $Ja$ is probably due to the increasing bubble rise velocity which 
limits their residence time in the cylinder.

\begin{figure}[t!]
\begin{center}
\includegraphics[width=8cm]{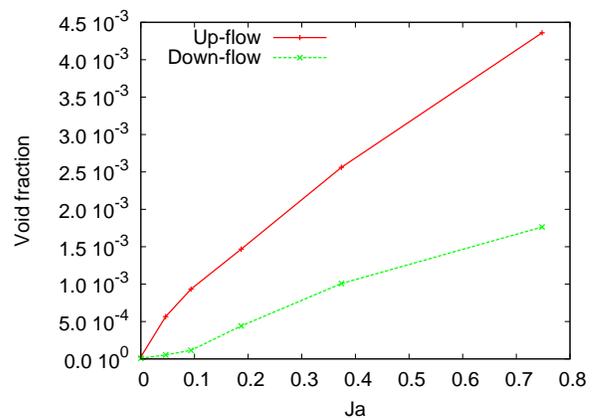}
\caption{Average void fraction in the up-flow and in the down-flow
regions for $N_b=5000$ bubbles.} 
\label{fig:voidN2}
\end{center}
\end{figure}

By calculating the volume of bubbles located in regions of positive and 
negative vertical liquid velocities we can look in detail at the 
effect of the increased buoyancy. Figure 
\ref{fig:voidN2} 
shows the time- and volume-averaged vapor volume fractions for
5,000 bubbles as a function of the Jacob 
number. The results for the other cases are similar, with smaller 
void fractions for 1,000 bubbles (for $Ja$ = 0.35, approximately 0.02\% and 
0.08\%), and larger ones for 10,000 bubbles (for $Ja$ = 0.35, approximately 
0.16\% and 0.36\%). 
It is seen that the void fraction in the upflow regions is 
consistently much larger than in the downflow regions, thus providing 
strong evidence for the expected destabilizing effect of the buoyancy 
provided by the bubbles. 

\begin{figure}
\begin{center}
\includegraphics[width=8cm]{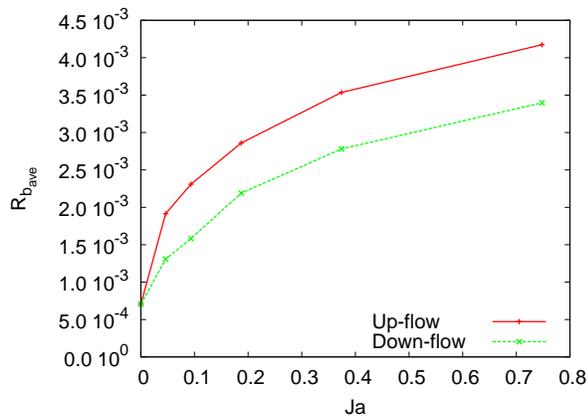}
\caption{Averaged radius of the bubble computed in the up-flow and
down-flow regions for $N_b=5000$ bubbles.}
\label{fig:RbN2}
\end{center}
\end{figure}

\begin{figure}
\begin{center}
\includegraphics[width=8cm]{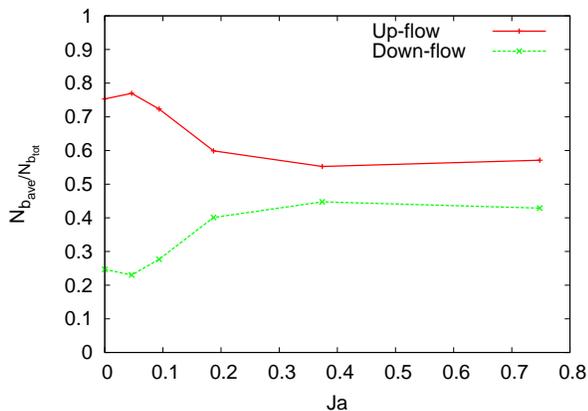}
\caption{Averaged bubble numbers in the up-flow and down-flow regions 
for $N_b=5000$ bubbles.}
\label{fig:NbN2}
\end{center}
\end{figure}

The void fraction reflects the combined effect of bubble number and bubble 
volume and it is interesting to consider these two contributions separately. 
The volume- and time-averaged bubble radius $\langle R_b\rangle_{V,t}$, 
defined by
\be
 \langle R_b\rangle_{V,t}\,=\,\left({3\over 4\pi N_b} \sum_i
\left< V_{bi}\right>_t
\right)^{1/3}
\ee
is shown in Fig. 
\ref{fig:RbN2} as a function of the Jacob number for the case of 
Fig.~\ref{fig:voidN2} with 5000 bubbles. As expected, the bubble size 
increases markedly with the Jacob number and  it tends to be somewhat 
larger in the hotter liquid regions. 
The time- and volume-averaged fractions of the total bubble number in the 
upflow and downflow regions, shown in Fig.~\ref{fig:NbN2}, indicates 
a strong tendency for bubbles to be in the hotter liquid regions, which 
is mostly responsible for the much larger void fraction in the 
rising liquid. This effect is probably due to fact that the newly injected 
bubbles at the hot plate tend to be swept up into the warm liquid by the 
convection current.

\begin{figure}
\begin{center} 
\includegraphics[width=8cm]{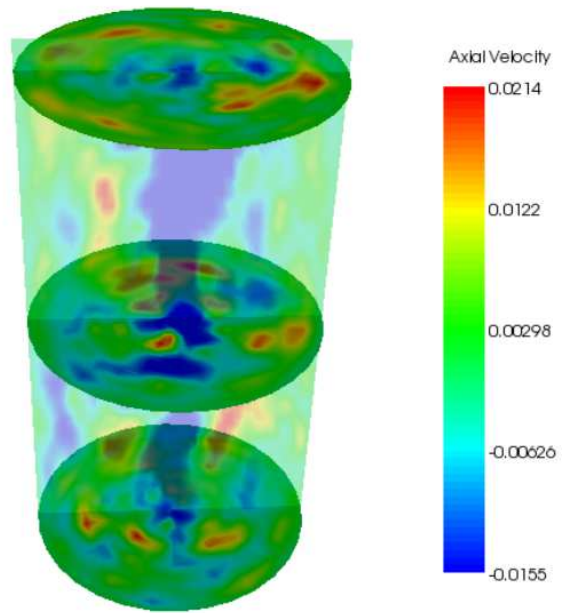}
\caption{Vertical and horizontal cross sections 
(taken at 0.05H, 0.5H, and 0.95H, respectively)
of the vertical liquid 
velocity distribution in the cylinder for $Ja=0$ and $N_b=5000$
 bubbles. The blue 
structure near the axis is the descending region of the toroidal vortex 
which prevails for small Jacob numbers.
The absolute values of the velocities are two order of magnitude
smaller as compared to the two subsequent figures as convection
is suppressed at $Ja=0$. 
}
\label{flstrJa0}
\end{center}
\end{figure}

The results of Fig.~\ref{fig:NbN2} for the bubble numbers show that 
the difference between the fractions of bubbles in the upflow and downflow 
regions is very large for small Jacob numbers and tends to decrease as $Ja$ 
increases. This behavior can be understood looking at the change in the 
flow structure. 

Without bubbles, the cylinder is occupied by a single convective roll 
which rises along one side and descends along the opposite side 
(Fig.~\ref{fig:snglphs}). 
A picture of the flow for the 5,000 bubbles, $Ja=0$ case is 
shown in Fig.~~\ref{flstrJa0} where one vertical and three horizontal cross 
sections color-coded with the vertical velocity field are displayed. The blue 
structure in the proximity of the cylinder axis is the descending region of a 
toroidal vortex, while the remaining green areas are those where the liquid 
rises, mostly with a smaller velocity, except for a few faster zones (yellow 
and red). It can be seen here that the volume occupied by the rising liquid is 
much greater than that occupied by the descending liquid, and this 
circumstance offers a likely explanation of the much smaller fraction of 
bubbles in the latter. 

\begin{figure}
\begin{center}
\includegraphics[width=8cm]{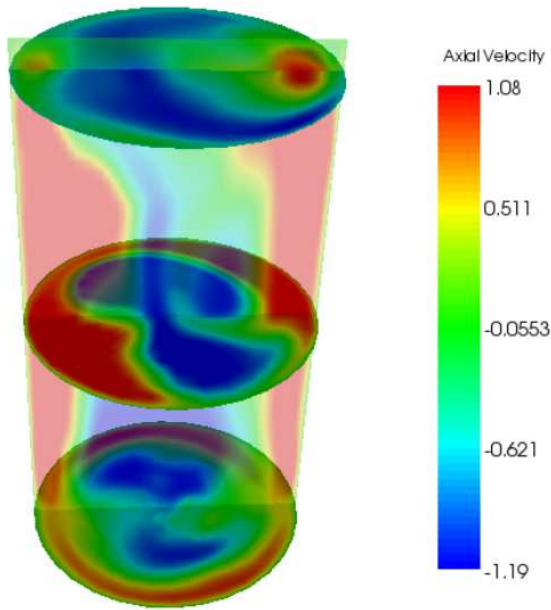}
\caption{Vertical and horizontal cross sections 
(taken at 0.05H, 0.5H, and 0.95H, respectively)
of the vertical liquid 
velocity distribution in the cylinder
 for $Ja=0.0935$ and 5,000 bubbles. The blue 
structure near the axis is the descending region of the toroidal vortex 
which prevails for small Jacob numbers.} 
\label{fig:Ja009}
\end{center}
\end{figure}

\begin{figure}
\begin{center}
\includegraphics[width=8cm]{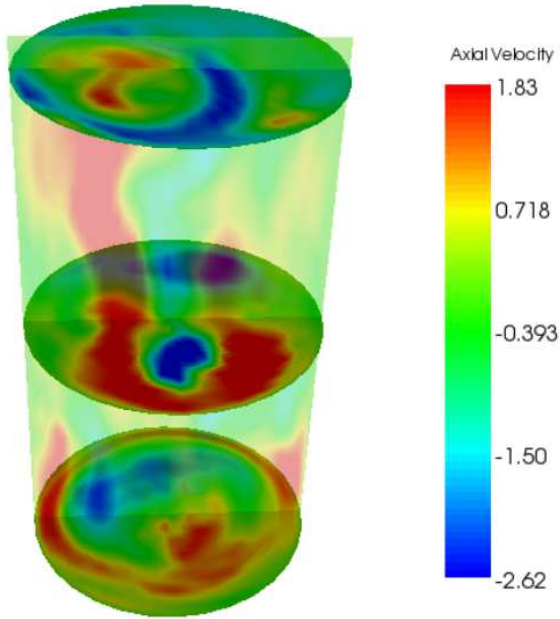}
\caption{Vertical and horizontal cross sections 
(taken at 0.05H, 0.5H, and 0.95H, respectively)
of the vertical liquid 
velocity distribution in the cylinder
 for $Ja=0.371$ and 5000 bubbles. The blue 
structure near the axis is the descending region of the toroidal vortex 
which prevails for small Jacob numbers.} 
\label{fig:Ja03}
\end{center}
\end{figure}

If the Jacob number is increased to  $Ja=0.0935$ (Fig.~~\ref{fig:Ja009}), 
the toroidal circulation is reinforced with a marked increase in the 
maximum rising and descending velocities (note that the color scales in 
these figures are not the same). For a still larger Jacob number, 
$Ja=0.374$ (Fig.~~\ref{fig:Ja03}) the flow has changed back to a circulation 
rising along one side of the cylinder and descending along the opposite one 
reminiscent of the single-phase pattern of Fig.~\ref{fig:snglphs}. 
Now the volumes occupied by the two streams are more balanced and the 
difference between the number of bubbles in the upflow and downflow 
regions is smaller, as seen in Fig.~ \ref{fig:NbN2}, although the 
bubble fraction in the upflow is still larger than in the downflow. 

\begin{figure}
\begin{center}
\includegraphics[width=8cm]{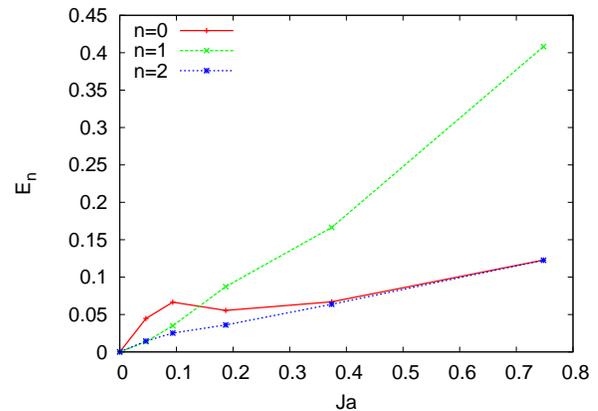}
\caption{Fourier modes of the kinetic energy in the angular direction 
for $N_b=5000$ bubbles. 
Mode 0 corresponds to a toroidal vortex and mode 1 to a circulatory motion 
in a vertical region with approximately horizontal axis.}
\label{fig:EneN2}
\end{center}
\end{figure}

\begin{figure}
\begin{center}
\includegraphics[width=8cm]{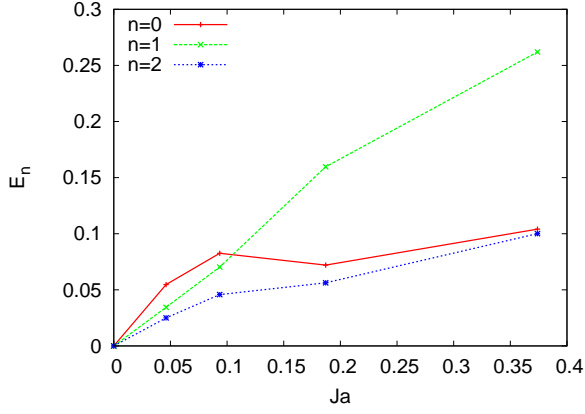}
\caption{Fourier modes of the kinetic energy in the angular direction 
for $N_b=10000$ bubbles. 
Mode 0 corresponds to a toroidal vortex and mode 1 to a circulatory motion 
in a vertical region with approximately horizontal axis.}
\label{fig:EneN3}
\end{center}
\end{figure}

These qualitative observations on the flow structure can be made quantitative 
by an analysis of the distribution of the liquid kinetic energy among 
different Fourier modes in the angular direction. We define the portion 
$E_n$ of the kinetic energy pertaining to mode $n$ by 
\be
  E_n\,=\, {\pi\over \beta g H^4 \Delta} \int_0^R r dr\int_0^H dz
\langle|{\bf u}_n|^2\rangle_t
\ee
where ${\bf u}_n$ is the $n$-th Fourier coefficient (in angular direction)
of the vector 
velocity field. 
%
%
The mode $n=0$ is axisymmetric and corresponds to a toroidal circulation 
symmetric about the 
vertical axis of the cylinder; $n=1$ is a single vortex around an approximately 
horizontal axis, and the higher modes give further information on the details 
of the distribution of the flow over the cross section of the cylinder. 
Results for the $n$ = 0, 1 and 2 modes are shown in Figs.\
\ref{fig:EneN2} and \ref{fig:EneN3} for 5,000 and 10,000 bubbles, 
respectively; the time averaging was carried out over the entire duration 
of the two-phase simulation. The values for $Ja=0$ are very small, but 
non-zero. It is seen here that, for zero or small Jacob 
number, most of the kinetic energy is in the toroidal mode $n$ = 0 which, 
for 5,000 bubbles, reaches a maximum at $Ja\simeq 0.09$, which is the case 
shown in Fig.~\ref{fig:Ja009}. For larger values of $Ja$, the energy in the 
$n=0$ mode decreases while that in the $n=1$ mode rapidly increases giving 
rise to the flow structure exemplified in Fig.~ \ref{fig:Ja03}.

\section{Summary and conclusions}\label{sec:sum}

In this paper we have presented a simple model to simulate the effect of 
phase change and two-phase flow on natural convection. While, for the 
reasons given in section~\ref{impl}, the results must be considered as 
preliminary, we have found that 
the addition of bubbles has a profound effect on the flow structure and 
on the Nusselt number. Bubbles that are prevented from growing by artificially 
maintaining the Jacob number equal to zero (corresponding to an 
infinitely large latent heat of vaporization) tend to short-circuit 
temperature non-uniformities and to stabilize the convective motion. As the 
Jacob number is increased, the added buoyancy due to the bubble growth 
rapidly increases the circulation and the heat transport. As the Jacob
number is increased further, bubble growth is rapid, the residence time 
short, and the rate of growth of the Nusselt number slows down. 
Correspondingly with the increasing Jacob number, the structure of the 
convective flow in the cylinder undergoes significant changes.

\vspace{0.5cm}

\noindent
{\it Acknowledgements:} The authors thank 
K.\ Sugiyama for various discussions and code validation 
 calculations. 
We moreover acknowledge 
SARA, Amsterdam,
 for supplying us with
CPU time.

\begin{appendix}

\section{Exact relations for the kinetic and thermal dissipations
$\epsilon_u$ and $\epsilon_\theta$}

Upon multiplying the momentum equation (\ref{momeq}) by {\bf u} 
and averaging over the cylinder volume and time, we find, by the no-slip condition 
on the cylinder walls,
\be
\epsilon_u \equiv \nu \langle \partial_j u_i \partial_j u_i \rangle_V =
\beta g \langle (T-T_{sat}) u_3 \rangle_V + {1\over \rho V} 
 \sum_n \left<{\bf f}_n\cdot{\bf u} \right>_t
\label{eps_u_int}
\ee
The term $\langle (T-T_{sat}) u_3 \rangle_V$ can be eliminated in terms 
of the single-phase Nusselt number at the hot base of the cylinder, given by 
(\ref{nubot}), to find 
\bea
 \epsilon_u &=& {\nu^3 \over H^4}{Ra \over Pr^2}\left(Nu_h -1\right)+ 
{1\over \rho V} 
 \sum_n \left<{\bf f}_{n,i}\cdot{\bf u} \right>_t \nonumber \\ &-&
{\beta g \over \rho c_p V}
\left< \sum_n \left(z_n-H\right)Q_n \right>_t
\eea 
in which $V=\pi R^2H$ is the volume of the cylinder. 
Alternatively, in terms of the Nusselt number at the cold top of the cylinder, 
\bea
 \epsilon_u &=& {\nu^3 \over H^4}{Ra \over Pr^2}\left(Nu_c -1\right)+ 
{1\over \rho V} 
 \sum_n \left<{\bf f}_{n,i}\cdot{\bf u} \right>_t \nonumber \\ &-&
{\beta g \over \rho c_p V} \left< \sum_n z_n Q_n \right>_t
\eea 

The thermal dissipation $\epsilon_\theta$ is defined in terms of 
\be
  \theta= T - {1\over 2} (T_h+T_c) \,=\,T-T_{sat}
\ee
as $\epsilon_\theta= \kappa \langle |\div \theta|^2\rangle_{V,t}$. 
An expression for this quantity may be readily obtained by 
multipling the energy equation by $\theta$ and averaging over the cylinder 
volume and time to find 
\bea
 \epsilon_\theta &=& {\kappa (T_h-T_c)\over 2H}\left[
-\langle \pat_3 T\rangle_{A,t,z=H} - \langle \pat_3 T\rangle_{A,t,z=0} 
\right] \nonumber \\ &+& {1\over \rho c_p} \left< \sum_n \theta_n Q_n\right>_t
\eea 
where we have used  the assumed insulation of the lateral walls and 
the fact that $\theta=\pm {1\over 2}(T_h-T_c)$ at the bottom and top of 
the cylinder. 
The temperature gradients can be eliminated in terms of the Nusselt 
numbers $Nu_{h,c} $ to find 
\be
 \epsilon_\theta= {\kappa \Delta^2\over H^2} {Nu_h +Nu_c  \over 2}
 + {1\over \rho c_p V} \left< \sum_n (T_n-T_{sat}) Q_n\right>_t
\ee 
which replaces the well-known relation 
$\epsilon_\theta= (\kappa \Delta^2/H^2) Nu$  of single-phase RB convection.

\end{appendix}


\begin{thebibliography}{10}

\bibitem{kad01}
L.~P. Kadanoff, Phys. Today {\bf 54},  34  (2001).

\bibitem{ahl09}
G. Ahlers, S. Grossmann, and D. Lohse, Rev. Mod. Phys. {\bf 81},  y  (2009).

\bibitem{dhi98}
V. Dhir, Ann. Rev. Fluid Mech. {\bf 30},  365  (1998).

\bibitem{MukherjeeDhir04}
A. Mukherjee and V. Dhir, J. Heat Transfer {\bf 126},  1023  (2004).

\bibitem{bun02a}
B. Bunner and G. Tryggvason, J. Fluid Mech. {\bf 466},  17  (2002).

\bibitem{bun02b}
B. Bunner and G. Tryggvason, J. Fluid Mech. {\bf 466},  53  (2002).

\bibitem{esm05}
A. Esmaeeli and G. Tryggvason, Phys. Fluids {\bf 17},  093303  (2005).

\bibitem{Elg}
S. Elghobaschi and T.~G. C., J. Fluid Mech. {\bf 242},  655  (1992).

\bibitem{Boi98}
M. Boivin, O. Simonin, and K. Squires, J. Fluid Mech. {\bf 375},  235  (1998).

\bibitem{FerranteElghobashi03}
A. Ferrante and S. Elghobashi, Phys. Fluids {\bf 15},  315  (2003).

\bibitem{cli99}
E. Climent and J. Magnaudet, Phys. Rev. Lett. {\bf 82},  4827  (1999).

\bibitem{maz03a}
I. Mazzitelli, D. Lohse, and F. Toschi, Phys. Fluids {\bf 15},  L5  (2003).

\bibitem{maz03b}
I. Mazzitelli, D. Lohse, and F. Toschi, J. Fluid Mech. {\bf 488},  283  (2003).

\bibitem{sug08b}
K. Sugiyama, E. Calzavarini, and D. Lohse, J. Fluid Mech. {\bf 608},  21
  (2008).

\bibitem{leg98b}
J. Magnaudet, J. Bor\'{e}e, and D. Legendre, Phys. Fluids {\bf 10},  1256
  (1998).

\bibitem{iva04}
O.~E. Ivashnyov and N.~N. Smirnov, Phys. Fluids {\bf 16},  809  (2004).

\bibitem{max94}
M.~R. Maxey, E. Chang, and L. Wang, Appl. Mech. Rev. {\bf 46},  6  (1994).

\bibitem{YangProsperetti08}
B. Yang and A. Prosperetti, J. Fluid Mech. {\bf 601},  253  (2008).

\bibitem{mag00}
J. Magnaudet and I. Eames, Annu. Rev. Fluid Mech. {\bf 32},  659  (2000).

\bibitem{nie07}
E.~A. van Nierop {\it et~al.}, J. Fluid Mech. {\bf 571},  439  (2007).

\bibitem{mei94}
R. Mei, J.~F. Klausner, and C.~J. Lawrence, Phys. Fluids {\bf 6},  418  (1994).

\bibitem{leg98}
D. Legendre and J. Magnaudet, J. Fluid Mech. {\bf 368},  81  (1998).

\bibitem{mag98}
J. Magnaudet and D. Legendre, Phys. Fluids {\bf 10},  550  (1998).

\bibitem{bat67}
G.~K. Batchelor, {\em An Introduction to Fluid Dynamics} (Cambridge University
  Press, Cambridge, 1967).

\bibitem{mag95}
J. Magnaudet, M. Rivero, and J. Fabre, J. Fluid Mech. {\bf 284},  97  (1995).

\bibitem{aut87}
T.~R. Auton, J. Fluid Mech. {\bf 183},  199  (1987).

\bibitem{riv91}
M. Rivero, J. Magnaudet, and J. Fabre, C.R. Acad. Sci. Paris II {\bf 312},
  1499  (1991).

\bibitem{cha95}
E.~J. Chang and M.~R. Maxey, J. Fluid Mech. {\bf 303},  133  (1995).

\bibitem{mei92}
R. Mei and J. Klausner, Phys. Fluids A {\bf 4},  63  (1992).

\bibitem{lab64}
D.~A. Labuntsov, B.~A. Kolchygin, E.~A. Zacharova, and L.~N. Vladimirova,
  Teplofiz. Vys. Temp. {\bf 2},  446  (1964).

\bibitem{ruc59}
E. Ruckenstein, Chem. Eng. Sci. {\bf 10},  22  (1959).

\bibitem{VeOr}
R. Verzicco and P. Orlandi, J. Comp Phys. {\bf 123},  402  (1996).

\bibitem{swartz}
P.~N. Swartztrauber, SIAM J. Numer. Anal. {\bf 11},  1136  (1974).

\bibitem{Verz3}
R. Verzicco and R. Camussi, J. Fluid Mech. {\bf 477},  19  (2003).

\bibitem{Shi79}
M. Shinohara and H. Hashimoto, J. Phys. Soc. Japan {\bf 46},  320  (1979).

\end{thebibliography}

\end{document}